\documentclass[conference]{IEEEtran}
\IEEEoverridecommandlockouts
\usepackage{cite}
\usepackage{amsmath,amssymb,amsfonts}
\usepackage{algorithmic}
\usepackage{graphicx}
\usepackage{textcomp}
\usepackage{mathtools}
\usepackage{xcolor}
\usepackage{siunitx}
\usepackage{float}
\usepackage{pgfplots}
\DeclareUnicodeCharacter{2212}{−}
\usepgfplotslibrary{groupplots,dateplot}
\usetikzlibrary{patterns,shapes.arrows}
\pgfplotsset{compat=newest}
\usepackage{standalone}
\usepackage{cite}
\usepackage{amsmath,amssymb,amsfonts}
\usepackage{algorithmic}
\usepackage{amsthm}
\usepackage{svg}
\usepackage{subcaption}
\usepackage{placeins}
\usepackage{standalone}
\usepackage{float}

\usetikzlibrary{spy}

\usepackage{siunitx}

\usepackage{xifthen}
\newcommand{\prob}[1][]{
\ifthenelse{\isempty{#1}}%
      {\ensuremath{P}}%
    {\ensuremath{P\left\(#1\right\)}}%
}
\newcommand{\vect}[1]{\boldsymbol{\mathrm{#1}}}
\newcommand{\mat}[1]{\boldsymbol{\mathrm{#1}}}

\newcommand{\diag}[1]{\mathrm{diag}\left(#1\right)}

\newcommand{\norm}[1]{\left\lVert#1\right\rVert}

\newcommand{\expt}[1]{\mathbb{E} \left\{#1\right\}}

\usepackage{tikz}
\usepackage{textcomp}
\usepackage{lipsum}

\usepackage[acronym,shortcuts]{glossaries}
\makeglossaries

\newacronym{3gpp}{3GPP}{3rd generation partnership project}
\newacronym{4g}{4G}{fourth-generation}
\newacronym{5g}{5G}{fifth-generation}
\newacronym{6g}{6G}{sixth-generation}
\newacronym{abp}{ABP}{authentication by personalisation}
\newacronym{aclr}{ACLR}{adjacent channel leakage ratio}
\newacronym{adc}{ADC}{analog-to-digital converter}
\newacronym{adr}{ADR}{adaptive data rate}
\newacronym{ag}{AG}{array gain}
\newacronym{ai}{AI}{artificial intelligence}
\newacronym{amam}{AM/AM}{amplitude modulation to amplitude modulation}
\newacronym{amp}{AMP}{approximate message passing}
\newacronym{ampm}{AM/PM}{amplitude modulation to phase modulation}
\newacronym{aoa}{AOA}{angle-of-arrival}
\newacronym{aod}{AOD}{angle-of-departure}
\newacronym{ap}{AP}{access point}
\newacronym{ar}{AR}{augmented reality}
\newacronym{arp}{ARP}{Antenna Reference Point}
\newacronym{asic}{ASIC}{application specific integrated circuit}
\newacronym{auv}{AUV}{autonomous underwater vehicle}
\newacronym{awgn}{AWGN}{additive white Gaussian noise}
\newacronym{bldc}{BLDC}{brushless DC}
\newacronym{bom}{BOM}{bill of materials}
\newacronym{bs}{BS}{base station}
\newacronym{bw}{BW}{bandwidth}
\newacronym{cad}{CAD}{channel activity detection}
\newacronym{cbm}{CBM}{condition based maintenance}
\newacronym{cc}{CC}{constant current}
\newacronym{ccdf}{CCDF}{complementary cumulative distribution function}
\newacronym{ccnn}{CCNN}{circular convolutional neural network}
\newacronym{cdf}{CDF}{cumulative distribution function}
\newacronym{cdrx}{CDRX}{connected mode DRX}
\newacronym{ce}{CE}{coverage enhancement}
\newacronym{ced}{CED}{cumulative energy density}
\newacronym{cf}{CF}{cell-free}
\newacronym{cfo}{CFO}{carrier frequency offset}
\newacronym{cots}{COTS}{commercial off-the-shelf}
\newacronym{cp}{CP}{cyclic prefix}
\newacronym{cpt}{CPT}{capacitive power transfer}
\newacronym{cpu}{CPU}{central-processing unit}
\newacronym{cqi}{CQI}{channel quality indicator}
\newacronym{cr}{CR}{coding rate}
\newacronym{crlb}{CRLB}{Cram\'er-Rao lower bound}
\newacronym{crs}{CRS}{cell reference signal}
\newacronym{cs}{CS}{compressed sensing}
\newacronym{csi}{CSI}{channel state information}
\newacronym{csp}{CSP}{contact service point}
\newacronym{css}{CSS}{chirp spread spectrum}
\newacronym{cv}{CV}{constant voltage}
\newacronym{dac}{DAC}{digital-to-analog converter}
\newacronym{daq}{DAQ}{data acquisition system}
\newacronym{dc}{DC}{direct current}
\newacronym{dcc}{DCC}{dynamic cooperation clustering}
\newacronym{de}{DE}{drain efficiency}
\newacronym{dl}{DL}{downlink}
\newacronym{dlc}{DLC}{Distributed Laser Charging}
\newacronym{dmimo}{D-MIMO}{distributed MIMO}
\newacronym{dpd}{DPD}{digital pre-distortion}
\newacronym{drx}{DRX}{Discontinuous Reception Mode}
\newacronym{e2e}{E2E}{end-to-end}
\newacronym{easa}{EASA}{European Union Aviation Safety Agency}
\newacronym{ecdf}{eCDF}{empirical cumulative distribution function}
\newacronym{ecsp}{ECSP}{edge computing service point}
\newacronym{edlc}{EDLC}{electrostatic double-layer capacitors}
\newacronym{edrx}{eDRX}{Extended Discontinuous Reception Mode}
\newacronym{ee}{EE}{energy efficiency}
\newacronym{egprs}{EGPRS}{Enhanced Data Rates for GSM Evolution}
\newacronym{eirp}{EIRP}{effective isotropic radiated power}
\newacronym{em}{EM}{electromagnetic}
\newacronym{embb}{eMBB}{enhanced Mobile Broadband}
\newacronym{en}{EN}{energy neutral}
\newacronym{eol}{EoL}{end of life}
\newacronym{epu}{EPU}{edge processing unit}
\newacronym{erp}{ERP}{effective radiated power}
\newacronym[plural=ESCs,firstplural=electronic speed controllers (ESCs)]{esc}{ESC}{electronic speed control}
\newacronym{etsi}{ETSI}{European Telecommunications Standards Institute}
\newacronym{evm}{EVM}{Error Vector Magnitude}
\newacronym{fa}{FA}{federation anchor}
\newacronym{fft}{FFT}{fast Fourier transform}
\newacronym{fh}{FH}{fronthaul}
\newacronym{fim}{FIM}{Fisher information matrix}
\newacronym{fom}{FoM}{figure of merit}
\newacronym{fpga}{FPGA}{field-programmable gate array}
\newacronym{fdma}{FDMA}{frequency-division multiple access}
\newacronym{gb}{GB}{grant-based}
\newacronym{gf}{GF}{grant-free}
\newacronym{gnb}{gNB}{Next Generation Node B}
\newacronym{gnn}{GNN}{graph neural network}
\newacronym{gnss}{GNSS}{global navigation satellite system}
\newacronym{gpclk}{GPCLK}{general purpose clock}
\newacronym{gprs}{GPRS}{General Packet Radio Services}
\newacronym{gps}{GPS}{Global Positioning System}
\newacronym{gpu}{GPU}{graphical processing unit}
\newacronym{gsm}{GSM}{Global System for Mobile Communications}
\newacronym{gwp}{GWP}{Global Warming Potential}
\newacronym{harq}{HARQ}{hybrid automatic repeat request}
\newacronym{hat}{HAT}{hardware attached on top}
\newacronym{hcs}{HCS}{human-centric services}
\newacronym{hpbm}{HPBM}{half power beam width}
\newacronym{HyMPRo}{HyMPRo}{Hybrid Multi-Path Routing algorithm}
\newacronym{ib}{IB}{in-band}
\newacronym{ibo}{IBO}{input back-off}
\newacronym{ic}{IC}{integrated circuit}
\newacronym{ict}{ICT}{information and communications technology}
\newacronym{id}{ID}{information decoding}
\newacronym{if}{IF}{intermediate frequency}
\newacronym{iid}{i.i.d.}{independently and identically distributed}
\newacronym{im}{IM}{intermodulation}
\newacronym{imu}{IMU}{inertial measurement unit}
\newacronym{iot}{IoT}{Internet of things}
\newacronym{ipt}{IPT}{inductive power transfer}
\newacronym{ipy}{IPY}{Interventions per Year}
\newacronym{ism}{ISM}{industrial, scientific and medical}
\newacronym{isp}{ISP}{internet service provider}
\newacronym{kkt}{KKT}{Karush--Kuhn--Tucker}
\newacronym{kpi}{KPI}{key performance indicator}
\newacronym{kvi}{KVI}{key value indicator}
\newacronym{lca}{LCA}{life cycle assessment}
\newacronym{lco}{LCO}{lithium cobalt oxide}
\newacronym{ldo}{LDO}{Low-dropout voltage regulator}
\newacronym{ldpc}{LDPC}{low-density parity-check}
\newacronym{lfp}{LFP}{lithium iron phosphate}
\newacronym{lic}{LIC}{lithium-ion capacitor}
\newacronym{liion}{Li-ion}{lithium-ion}
\newacronym{lipo}{LiPo}{lithium polymer}
\newacronym{lis}{LIS}{large intelligent surface}
\newacronym{llh}{LLH}{log-likelihood}
\newacronym{lmmse}{LMMSE}{least minimum mean square error}
\newacronym{lmo}{LMO}{lithium ion manganese oxide}
\newacronym{lora}{LoRa}{long range}
\newacronym{lorawan}{LoRaWAN}{long-range wide-area network}
\newacronym{los}{LoS}{line-of-sight}
\newacronym{lp}{LP}{linear programming}
\newacronym{lpt}{LPT}{laser power transfer}
\newacronym{lpwa}{LPWA}{Low Power Wide Area}
\newacronym{lpwan}{LPWAN}{low-power wide-area network}
\newacronym{lqi}{LQI}{link quality indicator}
\newacronym{lrelu}{LReLU}{leaky rectified linear unit}
\newacronym{lrt}{LRT}{likelihood-ratio test}
\newacronym{ls}{LS}{least squares}
\newacronym{lsfc}{LSFC}{large-scale fading component}
\newacronym{ltc}{LTC}{lithium thionyl chloride}
\newacronym{lte}{LTE}{long term evolution}
\newacronym{lto}{LTO}{lithium titanate}
\newacronym{m2m}{M2M}{machine to machine}
\newacronym{mac}{MAC}{Medium Access Control}
\newacronym{mcl}{MCL}{Maximum Coupling Loss}
\newacronym{mcs}{MCS}{modulation and coding scheme}
\newacronym{mcu}{MCU}{Microcontroller Unit}
\newacronym{mec}{MEC}{multi-access edge computing}
\newacronym{mems}{MEMS}{micro-electromechanical systems}
\newacronym{mf}{MF}{matched filter}
\newacronym{mimo}{MIMO}{multiple-input multiple-output}
\newacronym{miso}{MISO}{multiple-input single-output}
\newacronym{ml}{ML}{machine learning}
\newacronym{mlp}{MLP}{multilayer perceptron}
\newacronym{mmimo}{mMIMO}{massive MIMO}
\newacronym{mmse}{MMSE}{minimum mean square error}
\newacronym{mmtc}{mMTC}{massive machine-typed communication}
\newacronym{mmwave}{mmWave}{millimeter wave}
\newacronym{mpc}{MPC}{multipath component}
\newacronym{mppt}{MPPT}{maximum power point tracking}
\newacronym{mr}{MR}{maximum ratio}
\newacronym{mrc}{MRC}{maximum ratio combining}
\newacronym{mrc_em}{MRC}{maximum ratio combining}
\newacronym{mrt}{MRT}{maximum ratio transmission}
\newacronym{mse}{MSE}{mean square error}
\newacronym{mtc}{MTC}{Machine-Type Communication}
\newacronym{multi-rat}{Multi-RAT}{multiple radio access technology}
\newacronym{nb}{NB}{narrowband}
\newacronym{nbiot}{NB-IoT}{narrowband IoT}
\newacronym{nca}{NCA}{nickel cobalt aluminum}
\newacronym{nicd}{NiCd}{nikkel cadmium}
\newacronym{nimh}{NiMH}{nikkel metal hydride}
\newacronym{nlos}{NLoS}{non-line-of-sight}
\newacronym{nmc}{NMC}{nickel manganese cobalt}
\newacronym{nn}{NN}{neural network}
\newacronym{nnls}{NNLS}{non-negative least squares}
\newacronym{noma}{NOMA}{non-orthogonal multiple access}
\newacronym{nr}{NR}{New Radio}
\newacronym{ntp}{NTP}{network time protocol}
\newacronym{oai}{OAI}{OpenAirInterface} 
\newacronym{ofdm}{OFDM}{orthogonal frequency-division multiplexing}
\newacronym{ofdmim}{OFDM-IM}{OFDM with index modulation}
\newacronym{oob}{OOB}{out-of-band}
\newacronym{os}{OS}{operating system}
\newacronym{ota}{OTA}{over-the-air}
\newacronym{otaa}{OTAA}{over-the-air authentication}
\newacronym{p2p}{P2P}{point-to-point}
\newacronym{pa}{PA}{power amplifier}
\newacronym{pae}{PAE}{power-added efficiency}
\newacronym{papr}{PAPR}{peak-to-average power ratio}
\newacronym{pc}{PC}{pilot count}
\newacronym{pcb}{PCB}{printed circuit board}
\newacronym{pcg}{PCG}{power consumption gain}
\newacronym{pd}{PD}{powered device}
\newacronym{pdcch}{PDCCH}{physical downlink control channel}
\newacronym{pdp}{PDP}{power delay profile}
\newacronym{pdsch}{PDSCH}{physical downlink shared channel}
\newacronym{per}{PER}{packet error rate}
\newacronym{pg}{PG}{path gain}
\newacronym{pgd}{PGD}{proximal gradient descent}
\newacronym{pl}{PL}{path loss}
\newacronym{plr}{PLR}{packet loss ratio}
\newacronym{pll}{PLL}{phase-locked loop}
\newacronym{poe}{PoE}{power over Ethernet}
\newacronym{pps}{PPS}{pulse per second}
\newacronym{prbs}{PRBs}{Physical Resource Blocks}
\newacronym{psd}{PSD}{power spectral density}
\newacronym{pse}{PSE}{power sourcing equipment}
\newacronym{psm}{PSM}{power saving mode}
\newacronym{pss}{PSS}{primary synchronisation signal}
\newacronym{ptp}{PTP}{Precision Time Protocol}
\newacronym{pdr}{PDR}{packet delivery ratio}
\newacronym{ptw}{PTW}{paging time window}
\newacronym{qam}{QAM}{quadrature amplitude modulation}
\newacronym{qos}{QoS}{quality-of-service}
\newacronym{quadriga}{QuaDRiGa}{QUAsi Deterministic RadIo channel GenerAtor}
\newacronym{ra}{RA}{Random Access}
\newacronym{ran}{RAN}{radio access network}
\newacronym{rar}{RAR}{Random Access Response}
\newacronym{rat}{RAT}{radio access technology}
\newacronym{rbs}{RBS}{radio base station}
\newacronym{re}{RE}{radio element}
\newacronym[]{relu}{ReLU}{rectified linear unit}
\newacronym{rf}{RF}{radio frequency}
\newacronym{rfeh}{RFEH}{radio frequency energy harvesting}
\newacronym{rfic}{RFIC}{radio-frequency integrated circuit}
\newacronym{rfid}{RFID}{radio frequency identification}
\newacronym{rfpt}{RFPT}{radio frequency power transfer}
\newacronym{ris}{RIS}{reflective intelligent surface}
\newacronym{rms}{RMS}{root-mean-square}
\newacronym{rmse}{RMSE}{root-mean-square error}
\newacronym{ros}{ROS}{robot operating system}
\newacronym{rpi}{RPi}{Raspberry Pi}
\newacronym{rrc}{RRC}{Radio Resource Connection}
\newacronym{rreq}{RREQ}{route request packet}
\newacronym{rsrp}{RSRP}{Reference Signals Received Power}
\newacronym{rss}{RSS}{received signal strength}
\newacronym{rssi}{RSSI}{received signal strength indicator}
\newacronym{rtc}{RTC}{real time clock}
\newacronym{rtk}{RTK}{real time kinematics}
\newacronym{rts}{RTS}{ray tracing simulator}
\newacronym{rw}{RW}{RadioWeaves}
\newacronym{rx}{RX}{receiver}
\newacronym{rzf}{RZF}{regularized zero forcing}
\newacronym{sa}{SA}{synchronization anchor}
\newacronym{scfdma}{SCFDMA}{single-carrier frequency division multiple access}
\newacronym{sdg}{SDG}{Sustainable Development Goal}
\newacronym{sdn}{SDN}{software-defined network}
\newacronym{sdr}{SDR}{signal-to-distortion ratio}
\newacronym{sf}{SF}{spreading factor}
\newacronym{sfn}{SFN}{single frequency network}
\newacronym{sinr}{SINR}{signal-to-interference-plus-noise ratio}
\newacronym{siso}{SISO}{single-input single-output}
\newacronym{slam}{SLAM}{simultaneous localization and mapping}
\newacronym{slc}{SLC}{spatial leakage suppression}
\newacronym{smc}{SMC}{specular multipath component}
\newacronym{smps}{SMPS}{switched mode power supply}
\newacronym{sndr}{SNDR}{signal-to-noise-and-distortion ratio}
\newacronym{snidr}{SNIDR}{signal-to-noise-and-interference-and-distortion ratio}
\newacronym{snr}{SNR}{signal-to-noise ratio}
\newacronym{soc}{SoC}{state of charge}
\newacronym{ssb}{SSB}{synchronisation signal block}
\newacronym{ssd}{SSD}{solid state drive}
\newacronym{steam}{STEAM}{science, technology, engineering, the arts, and mathematics}
\newacronym{svd}{SVD}{singular value decomposition}
\newacronym{tau}{TAU}{tracking area update}
\newacronym{tcer}{TCER}{transported to consumed energy ratio}
\newacronym{tdd}{TDD}{time division duplexing}
\newacronym{tdoa}{TDOA}{time-difference-of-arrival}
\newacronym{toa}{TOA}{time-of-arrival}
\newacronym{tof}{ToF}{time-of-flight}
\newacronym{trp}{TRP}{Transmission Reception Point}
\newacronym{tsn}{TSN}{time-sensitive networking}
\newacronym{ttm}{TTM}{time to market}
\newacronym{ttn}{TTN}{The Things Network}
\newacronym{tx}{TX}{transmitter}
\newacronym{tsch}{TSCH}{time slotted channel hopping}
\newacronym{tdma}{TDMA}{time division multiple access}
\newacronym{uav}{UAV}{unmanned aerial vehicle}
\newacronym{udp}{UDP}{User Datagram Protocol}
\newacronym{ue}{UE}{user equipment}
\newacronym{ufpgd}{UfPGD}{unfolded proximal gradient descent}
\newacronym{ugv}{UGV}{unmanned ground vehicle}
\newacronym{uhd}{UHD}{USRP hardware driver}
\newacronym{uhf}{UHF}{ultra-high frequency}
\newacronym{ul}{UL}{uplink}
\newacronym{ula}{ULA}{uniform linear array}
\newacronym{upa}{UPA}{uniform planar array}
\newacronym{ura}{URA}{uniform rectangular array}
\newacronym{urllc}{URLLC}{ultra-reliable low-latency communications}
\newacronym{usrp}{USRP}{universal software radio peripheral}
\newacronym{uv}{UV}{unmanned vehicle}
\newacronym{uwb}{UWB}{ultrawideband}
\newacronym{vep}{VEP}{virtual edge platform}
\newacronym{vlc}{VLC}{visible light communication}
\newacronym{vlp}{VLP}{visible light positioning}
\newacronym{wb}{WB}{wideband}
\newacronym{wpt}{WPT}{wireless-power transfer}
\newacronym{wr}{WR}{white-rabbit}
\newacronym{wrsn}{WRSN}{wireless rechargeable sensor network}
\newacronym{wsn}{WSN}{wireless sensor network}
\newacronym{z3ro}{Z3RO}{zero third-order distortion}
\newacronym{zf}{ZF}{zero-forcing}
\newacronym{zmcscg}{ZMCSCG}{zero mean circularly symmetric complex Gaussian}

\def\BibTeX{{\rm B\kern-.05em{\sc i\kern-.025em b}\kern-.08em
    T\kern-.1667em\lower.7ex\hbox{E}\kern-.125emX}}
\begin{document}

\title{Deep Unfolding for Fast Linear Massive MIMO Precoders under a PA Consumption Model\\

\thanks{This work was made possible by a mobility grant provided by FWO (filenumber V424222N) and a short-term scientific mission grant by COST ACTION CA20120 INTERACT.  All code is available at: github.com/thomasf10/DeepUnfoldingMIMOPrecoder.\\
This paper is presented at VTC2023-Spring. T. Feys, X. Mestre, E. Peschiera, and F. Rottenberg, ``Deep Unfolding for Fast Linear Massive MIMO Precoders under a PA Consumption Model,'' in \textit{2023 IEEE 97th Vehicular Technology Conference (VTC2023-Spring)}, Florence, Italy, June 2023.}
}
\author{\IEEEauthorblockN{Thomas Feys*, Xavier Mestre\textsuperscript{\textdagger}, Emanuele Peschiera*, Fran\c{c}ois Rottenberg*
	}
	\IEEEauthorblockA{*KU Leuven, ESAT-WaveCore, Dramco, 9000 Ghent, Belgium
	}
    \IEEEauthorblockA{\textsuperscript{\textdagger}ISPIC, Centre Tecnologic Telecomunicacions Catalunya, Barcelona, Spain.}
}

\maketitle

\begin{abstract}
Massive \gls{mimo} precoders are typically designed by minimizing the transmit power subject to a \gls{qos} constraint. However, current sustainability goals incentivize more energy-efficient solutions and thus it is of paramount importance to minimize the consumed power directly. Minimizing the consumed power of the \gls{pa}, one of the most consuming components, gives rise to a convex, non-differentiable optimization problem, which has  been solved in the past using conventional convex solvers. Additionally, this problem can be solved using a \gls{pgd} algorithm, which suffers from slow convergence. In this work, in order to overcome the slow convergence, a deep unfolded version of the algorithm is proposed, which can achieve close-to-optimal solutions in only 20 iterations as compared to the 3500 plus iterations needed by the \gls{pgd} algorithm. Results indicate that the deep unfolding algorithm is three orders of magnitude faster than a conventional convex solver and four orders of magnitude faster than the \gls{pgd}.
\end{abstract}



\section{Introduction}

\subsection{Problem Formulation}
Reducing carbon emissions and energy consumption is more than ever a priority, as it is currently put forward by both Europe's Green Deal~\cite{greendeal} and the United Nations Sustainable Development Goals~\cite{sdgs}. Nevertheless, the estimated electricity consumption and carbon footprint of the wireless communications sector continues to rise~\cite{trends2040}. As such, when looking at the development of 6G, energy-reducing techniques are of the utmost importance. In current 5G networks, massive \gls{mimo} is one of the key enablers~\cite{mimo_reality}. Massive \gls{mimo} uses clever precoding schemes to spatially multiplex many users, leading to higher spectral efficiency~\cite{massivemimobook}. These precoders are typically obtained by minimizing the transmit power subject to a \gls{qos} constraint (e.g., a per-user \gls{sinr} constraint). However, it is known that the power consumed by the \glspl{pa}, one of the most consuming components, does not scale linearly with the transmit power~\cite{Grebennikov05, eff_model}. As such, minimizing the \glspl{pa} consumed power, rather than the transmit power, can lead to significant energy savings\cite{Peschiera22}. Moreover,~\cite{Cheng19} and \cite{Peschiera22} have shown, for single-user and multi-user systems respectively, that this typically leads to sparse solutions in the number of activated antennas. This can further be exploited by deactivating the RF chains of the unused antennas. Additionally, from~\cite{Cheng19,Peschiera22}, it is clear that minimizing the consumed power leads to a convex optimization problem, with the objective function being a composite of a convex, differentiable function and a convex, non-differentiable function. In~\cite{Peschiera22}, this problem is solved for a multi-user system using convex solvers. However, when aiming at real-time operation, faster algorithms are required. Given the form of the objective function, a \acrfull{pgd} algorithm can be used to find the global optimum. Unfortunately, \gls{pgd} suffers from slow convergence, requiring many iterations, which makes the use for real-time applications such as precoding difficult. Therefore, in this work, a deep unfolded version of the \gls{pgd} algorithm is proposed, which drastically speeds up the optimization, delivering close-to-optimal approximations, in only 20 iterations, as compared to the 5000 needed by \gls{pgd}.

\subsection{Deep Unfolding}
In recent years, algorithm unrolling, also known as deep unfolding, has provided many promising results in signal and image processing~\cite{eldar}. For instance, the technique has been successfully applied in compressed sensing, sparse coding, image denoising, detection and channel decoding in massive \gls{mimo} and many more application domains~\cite{eldar, lecun, unfolding_mimo}. Deep unfolding makes the connection between iterative algorithms and deep neural networks. In this scheme, each iteration of an iterative algorithm is represented as a layer in a neural network. This is done for a fixed number of layers/iterations to form a neural network of fixed size. Running the neural network emulates the execution of the traditional algorithm for a finite number of iterations. Furthermore, the main benefit over the traditional algorithm is the fact that algorithm-specific parameters are learned, using backpropagation and stochastic gradient descent-based optimization, while for the traditional algorithm these parameters need to be manually tuned. Learning these parameters in each layer/iteration, produces algorithms that are highly optimized for the problem at hand, and can as such provide faster convergence rates~\cite{lecun}. Many precoding problems lead to iterative optimization algorithms. Consequently, some works have proposed the use of deep unfolding to solve these problems. For instance, in~\cite{uf_sumrate_max} a deep unfolding algorithm is proposed to solve the sum-rate maximization problem in multi-user \gls{mimo}. Similarly, in~\cite{uf_sumrate_csierrors}, a deep unfolding algorithm is proposed that maximizes the sum-rate while being robust against channel estimation errors. Both works consider a transmit power constraint, neglecting the consumed power in their analysis. In this work, we consider the consumed power and minimized it, which results in a power consumption reduction.

\subsection{Contributions}
In this work, the design of a \gls{zf} precoder under a realistic \gls{pa} consumption model is studied. Hence, the \glspl{pa} consumed power is minimized, rather than the transmit power. First, we show that this leads to a convex optimization problem, with a composite objective function consisting of a convex, differentiable and a convex, non-differentiable part. Given the form of this cost function, the problem can be solved using a \gls{pgd} algorithm. To support real-time operations, a deep unfolded version of the \gls{pgd} algorithm is proposed. It is shown that this unfolded algorithm converges to close-to-optimal solutions, much faster than the traditional \gls{pgd}.

\textbf{Notations}: Vectors and matrices are denoted by bold lowercase and bold uppercase letters respectively. Superscripts $(\cdot)^*$, $(\cdot)^{\intercal}$ and $(\cdot)^{H}$ stand for the conjugate, transpose, and Hermitian transpose operators respectively. Subscripts $(\cdot)_m$ and $(\cdot)_k$ denote the antenna and user index. The expectation is denoted by $\expt{\cdot}$, while $\mat{D}_{\vect{a}}=\diag{\vect{a}}$ denotes a diagonal matrix whose diagonal entries are equal to the vector $\vect{a}$. The element at location $(i,j)$ in the matrix $\mat{A}$ is indicated as $[\mat{A}]_{i,j}$. The symbols $\norm{\cdot}_1$, $\norm{\cdot}_2$, $\norm{\cdot}_{2,1}$ and $\norm{\cdot}_F$ indicate the $L_1$, $L_2$, $L_{2,1}$ and Frobenius norms, respectively. $\mat{A}^{1/2}$ denotes the positive square root of $\mat{A}$, namely the only Hermitian positive semidefinite matrix such that $\mat{A}=\mat{A}^{1/2}\mat{A}^{1/2}$. 

\section{System Model}
\subsection{Signal Model}
Throughout this work, a massive \gls{mimo} \gls{bs} is considered, equipped with $M$ transmit antennas and serving $K$ single-antenna users. The precoded symbol vector $\vect{x}\in \mathbb{C}^{M\times 1}$ is given by
\begin{align}
    \vect{x} = \mat{W}^{\intercal}\vect{s}
\end{align}
where $\mat{W} \in \mathbb{C}^{K \times M}$ is the precoding matrix and $\vect{s}\in \mathbb{C}^{K \times 1}$ are the transmit symbols, which are assumed to be zero mean, uncorrelated and have unit power. The received signal $\vect{r} \in \mathbb{C}^{K\times 1}$ is then 
\begin{align}
    \vect{r} = \mat{H} \vect{x} + \vect{\nu}
\end{align}
where $\mat{H} \in \mathbb{C}^{K \times M}$ is the channel matrix. The vector $\vect{\nu} \in \mathbb{C}^{K\times 1}$ contains complex \gls{iid} \gls{awgn} with zero mean and variance $\sigma_\nu$.

\subsection{Power Consumption Model}
Given that the transmit symbols are uncorrelated, the transmit power at antenna $m$ is
\begin{align}
    p_m = \expt{|x_m|^2} = \sum_{k=0}^{K-1}|w_{k, m}|^2
\end{align}
where $x_m = [\vect{x}]_m$ is the precoded symbol at antenna $m$ and $w_{k, m} = [\mat{W}]_{k,m}$ is the precoding coefficient at antenna $m$ for user $k$. This gives a total transmit power 
\begin{align}\label{eq:txpower}
    p_{\mathrm{tx}} = \sum_{m=0}^{M-1}\sum_{k=0}^{K-1} |w_{k, m}|^2 = \norm{\mat{W}}_F^2.
\end{align}
When considering a realistic power consumption model, the transmit power does not scale linearly with the consumed power. In other words, the efficiency of the \gls{pa} is not fixed but varies with the transmit power, i.e., typically the closer the \gls{pa} operates to saturation, the higher the efficiency~\cite{eff_model}. For instance, for class B amplifiers, the efficiency scales with the square root of the transmit power~\cite{Grebennikov05, Cheng19}
\begin{align}
\eta_m = \eta_{\mathrm{max}} \sqrt{\frac{p_m}{p_{\mathrm{max}}}}
\end{align}
where $\eta_m$ is the efficiency of the $m$-th \gls{pa}, $\eta_{\mathrm{max}}$ is the maximal \gls{pa} efficiency and $p_{\mathrm{max}}$ is the maximal output power of the \gls{pa}.
The total consumed power by the \glspl{pa} is then 
\begin{equation}\label{eq:effmodel}
    \begin{aligned}
        p_{\mathrm{cons}} &= \sum_{m=0}^{M-1} \frac{p_m}{\eta_m} = \underbrace{\frac{\sqrt{p_{\mathrm{max}}}}{\eta_{\mathrm{max}}}}_{=\alpha} \sum_{m=0}^{M-1} \sqrt{p_m}\\
        &=\alpha \sum_{m=0}^{M-1} \sqrt{\sum_{k=1}^{K} |w_{k,m}|^2} =  \alpha \norm{\mat{W}}_{2,1}.
    \end{aligned}
\end{equation}

\section{Problem formulation}
Classical precoders such as \gls{zf} minimize the transmit power under a \gls{qos} constraint (e.g., a per-user \gls{sinr} constraint) \cite{massivemimobook}. However, when aiming to reduce the \glspl{pa}' consumed power, one should minimize the consumed power given in (\ref{eq:effmodel}) rather than the transmit power in (\ref{eq:txpower}). As such, a \gls{zf} precoder under the PA consumption model in (\ref{eq:effmodel}) can be formulated as
\begin{equation}\label{eq:optim1}
    \begin{aligned}
        \min_{\mat{W}} \quad &\alpha  \norm{\mat{W}}_{2,1}\\ 
         \mathrm{s.t.} \quad &\mat{H}\mat{W}^{\intercal} = \sigma_\nu\mat{D}_{\vect{\gamma}}^{1/2}
    \end{aligned}
\end{equation}
where $\mat{D}_{\vect{\gamma}} = \diag{\gamma_0,\dotsc,\gamma_{K-1}}$ and $\gamma_k$ is the target \gls{sinr} for user $k$.
In this formulation, the consumed power by the \glspl{pa} is minimized under a \gls{sinr} constraint per user. The precoder design problem (\ref{eq:optim1}) has been analyzed in \cite{Peschiera22} and can be solved by using convex optimization toolboxes (e.g., CVXPY \cite{diamond2016cvxpy}).
Problem (\ref{eq:optim1}) is equivalent to
\begin{equation}
    \begin{aligned}
    \min_{\mat{W}} \quad &\alpha  \norm{\mat{W}}_{2,1}\\ 
    \mathrm{s.t.} \quad & \norm{\mat{H}\mat{W}^{\intercal} - \sigma_\nu\mat{D}_{\vect{\gamma}}^{1/2}}_F^2 = 0.
    \end{aligned}
\end{equation}
This optimization problem can be reformulated using a Lagrangian function as follows
\begin{equation}
    \begin{split}
    \mathcal{L} &= \alpha  \left\|\mat{W}\right\|_{2,1} + \mu \norm{\mat{H}\mat{W}^{\intercal} - \sigma_\nu\mat{D}_{\vect{\gamma}}^{1/2}}_F^2 \\
    &= \mu\left(\lambda \left\| \mat{W}\right\|_{2,1} + \norm{\mat{H}\mat{W}^{\intercal} - \sigma_\nu\mat{D}_{\vect{\gamma}}^{1/2} }_F^2\right)
    \end{split}
    \label{eq:l}
\end{equation}
where $\lambda = \alpha/\mu$ and $\mu$ is the Lagrange multiplier.

\section{Proximal Gradient Descent}
The Lagrangian function in (\ref{eq:l}) is of the form 
\begin{align}
    \mathcal{L} = \lambda g(\mat{W}) + f(\mat{W})
\end{align}
with $f(\mat{W})$ a convex, differentiable function and $g(\mat{W})$ a convex, non-differentiable function. Given this form, the problem can be solved using \gls{pgd}~\cite{proxgd}. This algorithm first performs a gradient update with respect to the differentiable function $f(\mat{W})$. Next, the proximal operator of $\lambda g(\mat{W})$ is computed on the result of the gradient update. The general formulation of the iterative proximal algorithm is given by 
\begin{align}
    \mat{W}^{(i+1)} = \mathbf{prox}_{\lambda\eta g}\left(\mat{W}^{(i)} - \eta \nabla f(\mat{W}^{(i)}) \right)
\end{align}
where $\eta$ is the gradient step size and $\mathbf{prox}_{\lambda\eta g}(\cdot)$ is the proximal operator of the function $\lambda g(\cdot)$.
Since $\mat{W}$ is a complex matrix, $\nabla f(\mat{W})$ is defined as $\nabla f(\mat{W}) = \partial f(\mat{W})/\partial \mat{W}^* = \mat{W}\mat{H}^{\intercal} \mat{H}^* - \sigma_\nu\mat{D}_{\vect{\gamma}}^{1/2} \mat{H}^*$. Additionally, the proximal operator is only defined for real matrices, and consequently the real and imaginary parts of the complex matrix are concatenated in order to produce the following algorithm
\begin{equation}\label{eq:proxgd}
    \begin{aligned}
        &\mat{V} = \mat{W}^{(i)} - \eta \left(\mat{W}^{(i)}\mat{H}^{\intercal} \mat{H}^* - \sigma_\nu\mat{D}_{\vect{\gamma}}^{1/2} \mat{H}^*\right)  \\
        &\begin{bmatrix}
            \mat{W}_{\Re}^{(i+1)}\\
            \mat{W}_{\Im}^{(i+1)}
      \end{bmatrix} 
      = \mathbf{prox}_{\lambda \eta \norm{\cdot}_{2,1}}\left(
            \begin{bmatrix}
                \mat{V}_{\Re}\\
                \mat{V}_{\Im}
          \end{bmatrix}
          \right).
    \end{aligned}
\end{equation}

Here, $\lambda$ is dependent on the Lagrange multiplier $\mu$ and controls the trade-off between minimizing $g(\mat{W})$ and $f(\mat{W})$. Furthermore, convergence is guaranteed if the step size for the gradient update is $\eta \leq 1 / L$, where $L$ is the Lipschitz constant of $\nabla f(\mat{W}^{(i)})$, which is equivalent to an upper bound on the largest eigenvalue of $\mat{H}^{\intercal}\mat{H}^*$ \cite{proxgd}.

In order to define the proximal operator of the $L_{2,1}$ norm, we use the fact that the $L_{2,1}$ norm is a separable function, as it can be expressed as the $L_1$ norm over the column-wise $L_2$ norm. Hence, using the property of proximal operators for separable functions~\cite{prox_algo}, the proximal of the $L_{2,1}$ norm simply becomes a series of $L_2$-proximals on each column of $\mat{W}$. As such, the proximal operator for column $m$ of $\mat{W}$ (denoted as $\vect{w}_m$) can be expressed as

\begin{equation}
\begin{aligned}
    \mathbf{prox}_{\lambda \eta \norm{\cdot}_{2,1}}(\vect{w}_m) &= \left(1 - \frac{\lambda \eta}{\mathrm{max}(\norm{\vect{w}_m}_2, \lambda \eta)}\right) \vect{w}_m \quad \forall m \\
    &= \begin{cases}
        \left(1 - \frac{\lambda \eta}{\norm{\vect{w}_m}_2}\right) \vect{w}_m &\text{if $\norm{\vect{w}_m}_2 \geq \lambda \eta$}\\
        0 &\text{if  $\norm{\vect{w}_m}_2 < \lambda \eta $}
    \end{cases}.
\end{aligned}
\end{equation}

\begin{figure*}[!ht]
    \centering
     \begin{subfigure}[b]{.48\textwidth}
         \centering
	\includegraphics[width=0.8\linewidth]{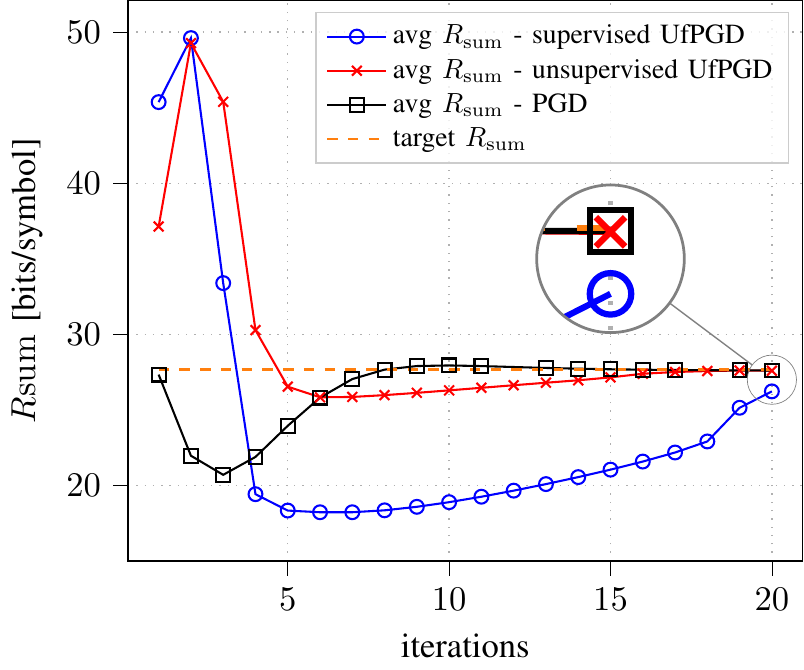}
  \caption{\footnotesize Average $R_{\mathrm{sum}}$}
  \label{fig:rsum}
     \end{subfigure}
     \begin{subfigure}[b]{.48\textwidth}
         \centering
	\includegraphics[width=0.8\linewidth]{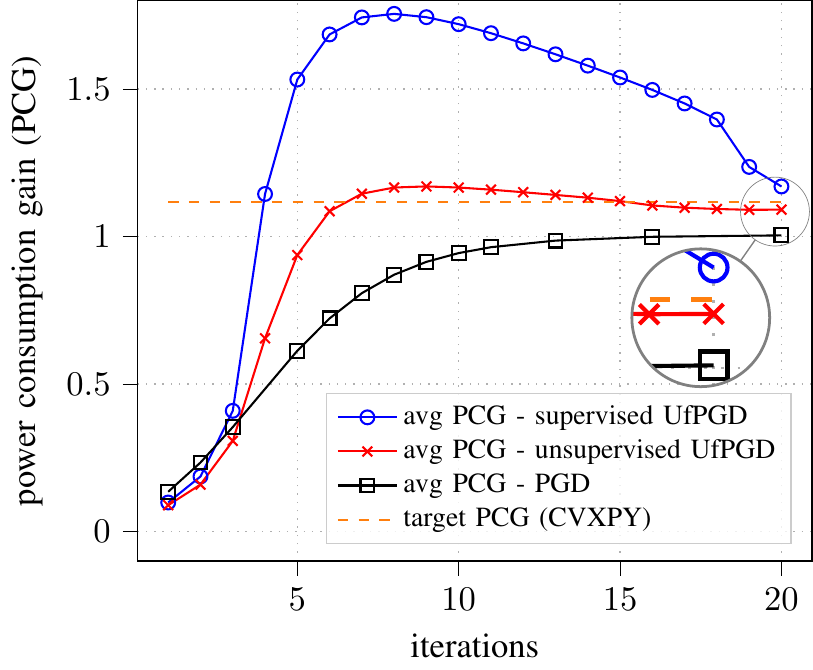}
  \caption{\footnotesize Average \acrlong{pcg}}
  \label{fig:pcg}
     \end{subfigure}
        \caption{\small Sum rate and \acrfull{pcg} convergence in function of the number of iterations/layers of \gls{pgd}, unfolding with supervised training and unfolding with unsupervised training. The results are averaged over 5000 channel realizations taken from the test set. Note that the target $R_{\mathrm{sum}}$ and \acrshort{pcg} (dashed) are obtained using CVXPY. }
        \label{fig:perf_vs_it}
\end{figure*}

\section{Deep Unfolded Proximal Gradient Descent}
The \gls{ufpgd} algorithm follows a similar structure as the algorithm in (\ref{eq:proxgd}). The algorithm is unrolled for a fixed number of iterations $I$. As such, a neural network consisting of $I$ layers is constructed, with each layer performing the following operations
\begin{equation} \label{eq:unfoldednet}
\begin{aligned}
    \mat{V} &= \mat{W}^{(i)} - \eta^{(i)} \left(\mat{W}^{(i)}\mat{H}^{\intercal} \mat{H}^* - \sigma_\nu\mat{D}_{\vect{\gamma}}^{1/2} \mat{H}^*\right) \\
    \begin{bmatrix}
        \mat{W}_{\Re}^{(i+1)}\\
        \mat{W}_{\Im}^{(i+1)}
  \end{bmatrix} 
  &= \mathbf{prox}_{\lambda^{(i)} \eta^{(i)} \norm{\cdot}_{2,1}}\left(
        \begin{bmatrix}
            \mat{V}_{\Re}\\
            \mat{V}_{\Im}
      \end{bmatrix}
      \right).
\end{aligned}
\end{equation}
The key difference between~(\ref{eq:unfoldednet}) and~(\ref{eq:proxgd}) is the fact that, in the unfolded iteration, the parameters $\lambda^{(i)}$ and $\eta^{(i)}$ are learned parameters that are optimized for each iteration/layer. More specifically, the parameters $\lambda^{(i)}$ and $\eta^{(i)}$ are learned by minimizing a certain cost function $C(\mat{W}^{(I)})$, where $\mat{W}^{(I)}$ is the output of the neural network (i.e., the result at the $I$-th layer). For instance, this can be done by using stochastic gradient descent in the following manner
\begin{align}
    \lambda^{(i)}_{\mathrm{new}} &= \lambda^{(i)} - \beta \frac{\partial C(\mat{W}^{(I)})}{\partial \lambda^{(i)} } \quad \forall i \in [1, 2, \cdots, I]\\
    \eta^{(i)}_{\mathrm{new}} &= \eta^{(i)} - \beta \frac{\partial C(\mat{W}^{(I)})}{\partial \eta^{(i)} } \quad \forall i \in [1, 2, \cdots, I]
\end{align}
where $\partial C(\mat{W}^{(I)})/\partial \eta^{(i)}$ and $\partial C(\mat{W}^{(I)})/\partial \lambda^{(i)}$ can be computed using backpropagation, and $\beta$ is the learning rate. The cost function can be defined in two ways, namely in a supervised or unsupervised manner. When training in a supervised way, the cost function is simply the \gls{mse} between the output of the neural network $\mat{W}^{(I)}$ and the known ground truth label for the optimal precoding matrix $\mat{W}_{\mathrm{gt}}$
\begin{align}
    C\left(\mat{W}^{(I)}\right) = \norm{\mat{W}^{(I)} - \mat{W}_{\mathrm{gt}}}_F^2.
\end{align}
The ground truth labels can be obtained using convex solvers such as CVXPY~\cite{diamond2016cvxpy}. 
When using the supervised learning approach, the \gls{mse} acts as a surrogate for the real cost function we want to optimize, which is the Lagrangian in (\ref{eq:l}). Conversely, one can also train the network in an unsupervised manner by replacing this surrogate loss function with the real cost function which is defined as
\begin{align}
    C\left(\mat{W}^{(I)}\right) = \lambda \left\| \mat{W}^{(I)}\right\|_{2,1} + \norm{\mat{H}\left(\mat{W}^{(I)}\right)^{\intercal} - \sigma_\nu\mat{D}_{\vect{\gamma}}^{1/2} }_F^2.
\end{align}
Finally, in order to ensure stability during the training, the values of $\lambda^{(i)}$ and $\eta^{(i)}$ are projected into their desired intervals. First, $\lambda^{(i)}$ should be in $[0, +\infty]$, leading to the following projection
$\lambda^{(i)}_{\mathrm{projected}} = \max(0, \lambda^{(i)})$.
Second, for the classical \gls{pgd} $\eta^{(i)}=1/L$, where $L$ is an upper bound on the largest eigenvalue of $\mat{H}^{\intercal}\mat{H}^*$. This upper bound can be approximated by the upper limit of the support of the Marchenko-Pastur law, which is (up to scaling) the asymptotic eigenvalue distribution for matrices with \gls{iid}, zero mean and unit variance entries. This approximation is given by $\tilde{L}=(\sqrt{K} + \sqrt{M})^2$~\cite{marchenkopastur}. This bound holds with probability one for sufficiently large $K$ and $M$. Hence, the value of $\eta^{(i)}$ is always projected into the following interval $[1/ (2\tilde{L}), 1/\tilde{L} ]$,
where the projection is defined as $\eta^{(i)}_{\mathrm{projected}} = \min ( \max (\eta^{(i)} , 1/ (2\tilde{L}) ), 1/\tilde{L} )$.

\section{Simulations}
\subsection{Training and Hyperparameters Selection}
In this section, simulations are performed to compare the performance of the unfolded and the classical \gls{pgd} algorithm. For the following simulations, the training set consists of $10^5$ Rayleigh fading channels sampled from a complex normal distribution with zero mean and variance one, i.e., $[\mat{H}]_{k, m}\sim \mathcal{CN}(0,1)$. The hyperparameters of the unfolding algorithm, such as the learning rate, batch size, etc. were tuned on a validation set of 5000 Rayleigh fading channels. Finally, all simulation results were obtained on an independent test set of 5000 Rayleigh fading channels. For training, the Adam optimizer~\cite{adam} is used with a batch size of 64 and a learning rate of 0.001. The unfolded algorithm consists of $I=20$ unfolded iterations/layers and was trained for 200 epochs, with early stopping if the validation loss did not further decrease. 

\subsection{Performance Evaluation Metrics}
To compare the performance of the unfolded and classical algorithms, two metrics are considered. First, the sum rate
\begin{align}
    R_{\mathrm{sum}} = \sum_{k=0}^{K-1} \mathrm{log}_2(1 + \mathrm{SINR}_k)
\end{align}
where $\mathrm{SINR}_k$ is the \gls{sinr} of user $k$. Given a per-user \gls{sinr} constraint of $10\ \mathrm{dB}$ and $K=8$ users, the target sum rate the algorithm should achieve is 27.68 bits/symbol. Second, the \gls{pcg} is defined as
\begin{align}
    \mathrm{PCG} = \frac{p_{\mathrm{cons}}^{\mathrm{ZF}}}{p_{\mathrm{cons}}^{\mathrm{ZF-eff}}}
\end{align}
where $p_{\mathrm{cons}}^{\mathrm{ZF}}$ is the consumed power of the \gls{zf} precoder, and $p_{\mathrm{cons}}^{\mathrm{ZF-eff}}$ is the consumed power of the proposed power efficient precoder (\ref{eq:optim1}), the consumed power is computed using (\ref{eq:effmodel}).

\subsection{Simulation Results}
To compare the convergence of \gls{pgd} and \gls{ufpgd}, the sum rate and \gls{pcg} are evaluated at each iteration/layer. This is done for $K=8, M=64$, a target per-user \gls{sinr} of $\gamma_k=10$ \SI{}{\decibel} and a noise variance of $\sigma_{\nu} = 1$. Note that $\lambda = 1/15$ for both the \gls{pgd} algorithm and the unsupervised cost function. Additionally, both \gls{pgd} and \gls{ufpgd} are initialized with the conjugate of the channel, i.e., $\mat{W}^{(0)} = \mat{H}^*$.

In Fig.~\ref{fig:rsum}, the sum rate per iteration is plotted for \gls{pgd} and \gls{ufpgd} with supervised and unsupervised training. From this figure it is clear that both \gls{pgd} and \gls{ufpgd} with unsupervised training can achieve the target sum rate of  27.68 bits/symbol. When training in a supervised manner, the unfolded algorithm is not able to achieve the goal sum rate. Given that for supervised training the \gls{mse} acts as a surrogate for the actual loss function, it is not able to capture the importance of achieving the \gls{sinr} constraint. When using unsupervised training, the real objective function is optimized, which enables the algorithm to achieve the \gls{sinr} constraint.

\begin{figure}[!t]
    \centering
    \includegraphics[width=.78\linewidth]{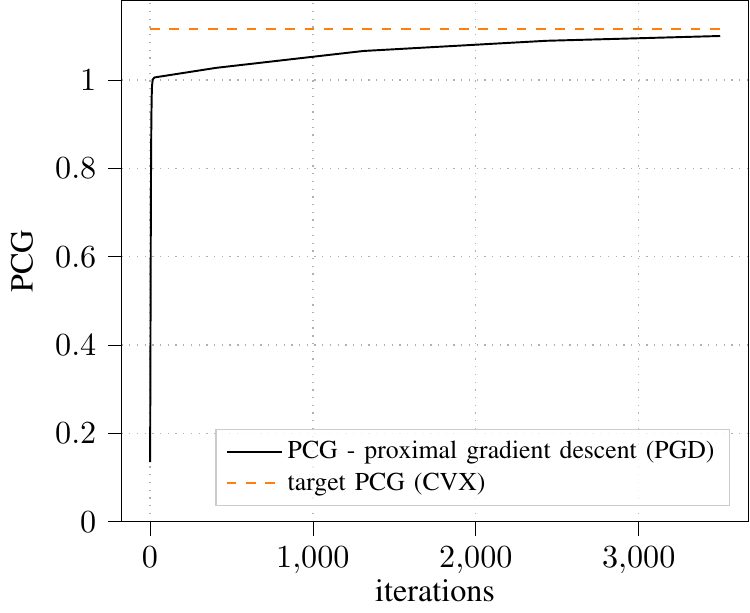}
    \caption{\small \Acrfull{pcg} per iteration of \gls{pgd}, versus the target \acrshort{pcg} of CVXPY. Averaged over 5000 channel realizations. }
    \label{fig:proxgd}
\end{figure}

In Fig.~\ref{fig:pcg}, the \gls{pcg} per iteration is plotted for \gls{pgd} and \gls{ufpgd}. This figure shows that after 20 iterations, the \gls{pgd} algorithm reaches a \gls{pcg} of approximately one, indicating that it consumes the same amount of energy as the classical \gls{zf} precoder. It is widely known that \gls{pgd} suffers from slow convergence, as is illustrated in Fig.~\ref{fig:proxgd}, where the \gls{pcg} is depicted when the algorithm is run for more iterations. Indeed, from this figure, it is clear that \gls{pgd} does eventually reach the target \gls{pcg}. However, it needs over 3500 iterations to get close to the optimal solution. This highlights the benefit of using the unfolded algorithm, as after 20 iterations it is close to the optimal solution. This is depicted in Fig.~\ref{fig:pcg}, where after 20 iterations the unsupervised \gls{ufpgd} algorithm reaches a $\mathrm{PCG} = 1.0915$, where the target value (obtained with CVXPY) is $\mathrm{PCG} = 1.1160$. 

\begin{figure}[!t]
    \centering
    \includegraphics[width=.8\linewidth]{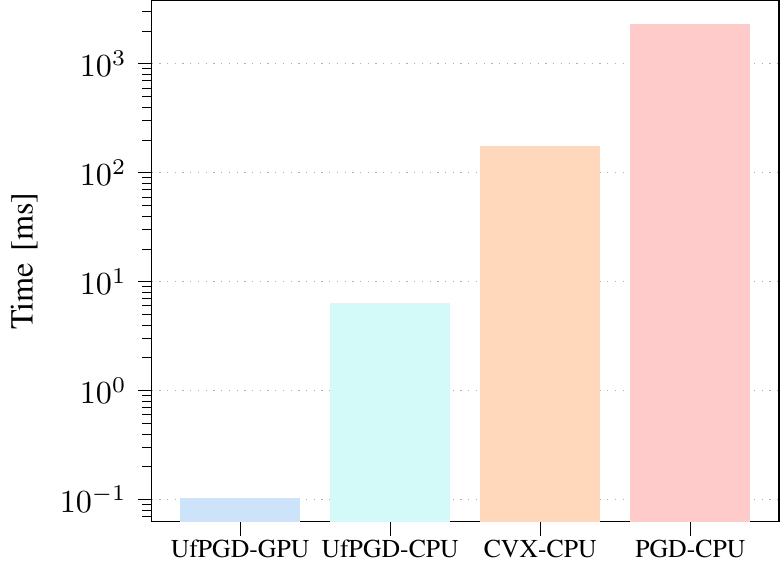}
    \caption{\small Execution time of \gls{ufpgd} on GPU and CPU, CVX and \gls{pgd}, both on CPU (ran for 5000 iterations per execution). Averaged over 5000 executions.}
    \label{fig:speed}
\end{figure}

When comparing the execution time of \gls{ufpgd}, \gls{pgd}\footnote{In order to reach the optimal solution, \gls{pgd} is executed for 5000 iterations.} and CVXPY in Fig.~\ref{fig:speed}, it is clear that the unfolding algorithm has the smallest execution time. This is expected as the unfolding algorithm only needs 20 iterations compared to the 5000 needed by \gls{pgd}. Additionally, it is clear that general-purpose numerical solvers such as CVXPY cannot reach sufficiently low execution times, as the execution times are three orders of magnitude larger than the unfolding algorithm.

\section{Conclusion}
In this work, a fast unfolded optimization algorithm is developed, for a massive MIMO precoder that minimizes the consumed power rather than the transmit power. It is shown that the optimization problem gives rise to an objective function that consists of a convex, differentiable and a convex, non-differentiable part. Given this form, the optimization problem is solved using \gls{pgd}. However, given the slow convergence rate of the \gls{pgd} algorithm, a deep unfolded version of the algorithm is proposed. The proposed algorithm is highly optimized for the task at hand, which allows it to achieve close-to-optimal solutions in only 20 iterations, as compared to the 3500 plus iterations needed by the \gls{pgd} algorithm. Finally, the execution time of the proposed algorithm is compared against a conventional numerical convex solver (CVXPY), showing that the proposed algorithm is three orders of magnitude faster.

\bibliographystyle{IEEEtran}
\bibliography{IEEEabrv,mybib}

\begin{thebibliography}{10}
\providecommand{\url}[1]{#1}
\csname url@samestyle\endcsname
\providecommand{\newblock}{\relax}
\providecommand{\bibinfo}[2]{#2}
\providecommand{\BIBentrySTDinterwordspacing}{\spaceskip=0pt\relax}
\providecommand{\BIBentryALTinterwordstretchfactor}{4}
\providecommand{\BIBentryALTinterwordspacing}{\spaceskip=\fontdimen2\font plus
\BIBentryALTinterwordstretchfactor\fontdimen3\font minus
  \fontdimen4\font\relax}
\providecommand{\BIBforeignlanguage}[2]{{%
\expandafter\ifx\csname l@#1\endcsname\relax
\typeout{** WARNING: IEEEtran.bst: No hyphenation pattern has been}%
\typeout{** loaded for the language `#1'. Using the pattern for}%
\typeout{** the default language instead.}%
\else
\language=\csname l@#1\endcsname
\fi
#2}}
\providecommand{\BIBdecl}{\relax}
\BIBdecl

\bibitem{greendeal}
{European Commission}, ``{The European Green Deal},'' \emph{COM (2019)},
  November 2019.

\bibitem{sdgs}
{United Nations}, ``{The 2030 Agenda and the Sustainable Development Goals: An
  opportunity for Latin America and the Caribbean},'' {(LC/G.2681-P/Rev.3),
  Santiago, 2018}.

\bibitem{trends2040}
L.~Belkhir and A.~Elmeligi, ``\BIBforeignlanguage{en}{Assessing {ICT} global
  emissions footprint: {Trends} to 2040 \& recommendations},''
  \emph{\BIBforeignlanguage{en}{Journal of Cleaner Production}}, vol. 177, pp.
  448--463, Mar. 2018.

\bibitem{mimo_reality}
E.~Björnson, L.~Sanguinetti, H.~Wymeersch, J.~Hoydis, and T.~L. Marzetta,
  ``Massive mimo is a reality—what is next?: Five promising research
  directions for antenna arrays,'' \emph{Digital Signal Processing}, vol.~94,
  pp. 3--20, 2019, special Issue on Source Localization in Massive MIMO.

\bibitem{massivemimobook}
E.~Bj\"{o}rnson, J.~Hoydis, and L.~Sanguinetti, ``Massive {MIMO} networks:
  {Spectral}, energy, and hardware efficiency,'' \emph{Foundations and
  Trends{\textregistered} in Signal Processing}, vol.~11, no. 3-4, pp.
  154--655, 2017.

\bibitem{Grebennikov05}
A.~Grebennikov, \emph{{RF and Microwave Power Amplifier Design}}, 1st~ed.\hskip
  1em plus 0.5em minus 0.4em\relax NY, USA: McGraw-Hill, 2005.

\bibitem{eff_model}
S.~Mikami, T.~Takeuchi, H.~Kawaguchi, C.~Ohta, and M.~Yoshimoto, ``{An
  Efficiency Degradation Model of Power Amplifier and the Impact against
  Transmission Power Control for Wireless Sensor Networks},'' in \emph{2007
  IEEE Radio and Wireless Symposium}, 2007, pp. 447--450.

\bibitem{Peschiera22}
E.~Peschiera and F.~Rottenberg, ``{Linear Precoder Design in Massive MIMO under
  Realistic Power Amplifier Consumption Constraint},'' in \emph{42nd WIC
  Symposium on Information Theory and Signal Processing in the Benelux}, 2022.

\bibitem{Cheng19}
H.~V. Cheng, D.~Persson, and E.~G. Larsson, ``{Optimal MIMO Precoding Under a
  Constraint on the Amplifier Power Consumption},'' \emph{IEEE Transactions on
  Communications}, vol.~67, no.~1, pp. 218--229, 2019.

\bibitem{eldar}
V.~Monga, Y.~Li, and Y.~C. Eldar, ``Algorithm unrolling: Interpretable,
  efficient deep learning for signal and image processing,'' \emph{IEEE Signal
  Processing Magazine}, vol.~38, no.~2, pp. 18--44, 2021.

\bibitem{lecun}
K.~Gregor and Y.~LeCun, ``Learning fast approximations of sparse coding,'' in
  \emph{Proceedings of the 27th International Conference on International
  Conference on Machine Learning}, ser. ICML'10.\hskip 1em plus 0.5em minus
  0.4em\relax Madison, WI, USA: Omnipress, 2010, p. 399–406.

\bibitem{unfolding_mimo}
\BIBentryALTinterwordspacing
A.~Balatsoukas-Stimming and C.~Studer, ``Deep unfolding for communications
  systems: A survey and some new directions,'' 2019. [Online]. Available:
  \url{https://arxiv.org/abs/1906.05774}
\BIBentrySTDinterwordspacing

\bibitem{uf_sumrate_max}
Q.~Hu, Y.~Cai, Q.~Shi, K.~Xu, G.~Yu, and Z.~Ding, ``Iterative algorithm induced
  deep-unfolding neural networks: Precoding design for multiuser mimo
  systems,'' \emph{IEEE Transactions on Wireless Communications}, vol.~20,
  no.~2, pp. 1394--1410, 2021.

\bibitem{uf_sumrate_csierrors}
J.~Xu, C.~Kang, J.~Xue, and Y.~Zhang, ``A fast deep unfolding learning
  framework for robust mu-mimo downlink precoding,'' \emph{IEEE Transactions on
  Cognitive Communications and Networking}, pp. 1--1, 2023.

\bibitem{diamond2016cvxpy}
S.~Diamond and S.~Boyd, ``{CVXPY}: {A} {P}ython-embedded modeling language for
  convex optimization,'' \emph{J. Mach. Learn. Res.}, vol.~17, no.~83, pp.
  1--5, 2016.

\bibitem{proxgd}
\BIBentryALTinterwordspacing
F.~Bach, R.~Jenatton, J.~Mairal, and G.~Obozinski, ``{Optimization with
  Sparsity-Inducing Penalties},'' 2011. [Online]. Available:
  \url{https://arxiv.org/abs/1108.0775}
\BIBentrySTDinterwordspacing

\bibitem{prox_algo}
\BIBentryALTinterwordspacing
N.~Parikh and S.~Boyd, ``{Proximal Algorithms},'' \emph{Found. Trends Optim.},
  vol.~1, no.~3, p. 127–239, jan 2014. [Online]. Available:
  \url{https://doi.org/10.1561/2400000003}
\BIBentrySTDinterwordspacing

\bibitem{marchenkopastur}
V.~A. Mar{\~c}enko and L.~A. Pastur, ``Distribution of eigenvalues for some
  sets of random matrices,'' \emph{Mathematics of The Ussr-sbornik}, vol.~1,
  pp. 457--483, 1967.

\bibitem{adam}
\BIBentryALTinterwordspacing
D.~P. Kingma and J.~Ba, ``Adam: A method for stochastic optimization,'' 2014.
  [Online]. Available: \url{https://arxiv.org/abs/1412.6980}
\BIBentrySTDinterwordspacing

\end{thebibliography}

\end{document}